# Error Correction and Digitalization Concepts in Biochemical Computing

**Leonid Fedichkin**, **Evgeny Katz** and **Vladimir Privman**\*

Department of Chemistry and Biomolecular Science, and
Department of Physics, Clarkson University,
Potsdam, New York 13699, USA

**Key Words:** biochemical computing, error correction, fanout, logic gate, XOR, NAND, rectification

**Abstract**

We offer a theoretical design of new systems that show promise for digital biochemical computing, including realizations of error correction by utilizing redundancy, as well as signal rectification. The approach includes information processing using encoded DNA sequences, DNAzyme biocatalyzed reactions and the use of DNA-functionalized magnetic nanoparticles. Digital XOR and NAND logic gates and copying (fanout) are designed using the same components.

## 1. Introduction

Biochemical computing is an exciting new field which shows great promise, but at the same time faces substantial challenges. Recently, research efforts were directed at unconventional chemical computing[1–3] based on a chemical "soup" where data are represented by varying concentrations of chemicals.[4,5] The computations are performed by naturally occurring chemical reactions that can proceed in a solution or at an interface functionalized with catalytic sites. The algorithmic problems can be solved at the level of a single molecule,[6] resulting in dramatic miniaturization and allowing parallel computation performed by numerous molecules.[7] So far, the field is still in a very early experimental stage, but may have great future potential.[8] Biochemical computing (biocomputing) is a special kind of chemical computing based on enzyme-catalyzed reactions[9–12] and DNA recognition/reaction processes.[13–18] The DNA based processes can include hybridization of complementary strands as well as enzyme-catalyzed DNA reactions such as scission, replication, etc. Although the ultimate scalability of the DNA computing, in its present variants, could be questioned, it shows promise to address certain complex combinatorial problems.[19–21] Moreover, the recent discovery[22] of catalytic DNAzymes allows the computing processes typical for enzyme-catalyzed reactions to be catalyzed by DNA molecules. The biocomputing systems can be based on the existing chip technologies currently used for biosensing (particularly DNA chips).[23] Further scaling down of biocomputing (particularly DNA computing) devices to nano-size is possible.[24] Substantial research efforts were reported on the development of special "molecular" software to operate with biocomputers.[25]

Biocomputing also poses interesting theoretical challenges. On the conceptual level, it might help us understand how living organisms manage to control extremely complex and coupled biochemical reactions, by casting the biochemical (metabolic) pathways in living organisms in the language of information theory. Even without full scalability, biochemical computing, especially at biomolecule-functionalized interfaces[26,27] rather than in the bulk, shows promise of providing the mechanisms to better couple ordinary electronics with biological organism signaling. Finally, to achieve complexity on par with ordinary electronics, biocomputing should be researched for ways to minimize/correct errors and develop "digitalization" concepts, as further addressed in Section 2.

In this work, we introduce new systems and ideas, detailed in Sections 3-5, that show *promise for digital biochemical computing and error correction*. In order to introduce the notation, the rest of this section is devoted to a short overview of selected concepts and challenges in biochemical computing.

The experimental activity in the field of chemical computing has been mostly directed towards the development of chemical systems operating as logic gates.[28–30] Various molecular- and supra-molecular systems were developed to mimic different logic operations.[31–33] Chemical or/and physical input signals were used to activate chemical logic gates. For example, molecular assemblies activated by chemical and light inputs were reported as logic AND gates with the optical readout of the output signals,[34,35] while a quinone-monolayer assembled on an electrode was used as a pH/light activated AND gate with the electrochemical readout of the output signals.[36] Molecule-based XOR or InhibA logic gates were also constructed.[37,38]

Chemical systems with simultaneous parallel operation of several different gates allow simple arithmetic operations. Specifically, chemical systems exhibiting XOR and AND logic gate functions were used to realize a half-adder,[39–41] while other chemical systems that performed InhibA and XOR operations were used to make a half-subtractor.[42,43] However, most results reported for the chemical logic systems represent the most trivial AND or OR logic gates. In order to assemble more complex logic gates or their combinations, very complex supra-molecular chemical systems were used. All constructed chemical gates and computing units use different input signals to activate, and therefore it is difficult to assemble them into a composite "device" with standard input/output signals. The biggest problem with chemical logic systems is the low specificity of most chemical reactions. This does not allow several different reactions in one system to run without interference between them. The "cross-talking" between the different reactions hinders assembling multi-component/multi-functional chemical logic systems and computing units.

To a large extent, this problem can be solved by means of *biocomputing* using *enzyme-based logic systems*.[9–11] The enzymes provide high specificity for the running reactions without interference between the reacting compounds and without "cross-talking" between the logic gates. Thus, several enzymes working in parallel in the same homogeneous system were able to perform concerted operations yielding simple computing units: half-adder and half-subtractor.[9–11] Several logic gates (AND, OR, XOR, NOR, InhibA, InhibB, etc.) were constructed with *standard input signals* (specifically additions of $H_2O_2$ as input A, and glucose as input B).[11] For example, an AND gate composed of two enzymes, glucose oxidase (GOx) and catalase (Cat), operated upon the application of two standard input signals: the addition of $H_2O_2$ and glucose.[11] The output signal was defined as the absorbance generated by the biocatalytically produced gluconic acid (measured in the presence of hydroxylamine and $Fe^{3+}$). The TRUE signal ("1") corresponding to the formation of gluconic acid was observed only when GOx was activated in the presence of glucose and oxygen, while oxygen was biocatalytically produced by Cat only upon addition of $H_2O_2$. Separate additions of glucose (in the absence of $H_2O_2$) or $H_2O_2$ (in the absence of glucose) did not result in any absorbance change in the system (FALSE signal: "0"). Thus, the behavior of the system represented the AND logic gate.

The other enzyme-based logic gates operate in a similar manner.[9–11] The large variety and high specificity of available enzymes make assembling different logic gates into a network possible. A network composed of sequential and parallel logic gates could be assembled from various enzymes modeling natural metabolic pathways. For example, a sequence of concatenated logic gates composed of four enzymes (acetylcholine esterase, choline oxidase, microperoxidase-11, and the $NAD^+$-dependent glucose dehydrogenase) performing sequentially OR, AND, and XOR logic operations was constructed to prove the concept of the *enzyme-based computing networks*.[12] This system accepts four chemical input signals (additions of acetylcholine, butylcholine, oxygen, and glucose in 16 different combinations), producing one output signal (concentration of NADH) that is dependent on all input signals. Similarly to the enzyme-based concatenated gates, DNA-based NOR and OR logic gates were integrated into logic circuits.[44] Logic circuitries of higher complexity, integrating 128 DNA-based logic gates, 32 input DNA molecules, and 8 two-channel fluorescent outputs across 8 wells, were also demonstrated recently.[45] Biocatalytic systems based on enzymes can generate signals using self-produced power, while operating as biofuel cells,[46] thus resulting in self-powered logic gates.

## 2. Scalability, Error Correction, and Digitalization Concepts

Logic networks composed of enzymes and/or DNA working in concert allow many chemical input signals, and provide output response(s) based upon the performance of all logic operations in the system. Enzyme/DNA-based logic systems and biocomputing units of rather high complexity could be envisaged, and numerous applications could be expected. Error-reduction/correction in biocomputing systems[47,48] becomes extremely important as the complexity of numerous connected biochemical reactions increases. This is needed for scalability and fault-tolerance of the multi-gate computation process. Among the established error-correction paradigms, making the information processing digital is perhaps the most important. For both analog and digital information processing, in practice error suppression[49,50] can be accomplished **(i)** by nonlinearity of the "filters" though which the signals are passed, **(ii)** by feedback, and **(iii)** by redundancy, as well as by combinations of these techniques. The third approach is widely used especially for digital information processing,[49] whereas the first two approaches are more suited for analog information.

Nonlinear signal filtering is not easy to accomplish in complex biochemical systems: it would imply some nonlinear (in the input concentrations) manipulation of chemical compositions. This is not generally available, though we do offer a possible "signal rectification" approach for our system in Section 5. Similarly, most of the presently studied few-component (single or few logical gates) systems are unlikely to offer possibilities for controlled utilization of feedback in a complicated sequence of chemical reactions, in which some of the output products would also need to play the role of input reactants for "earlier" gates. Therefore, as the first step we suggest considering redundancy as the error suppression mechanism: in the next section we will propose a paradigm of biochemical systems which advances us towards *digital* information processing, while allowing the redundancy (copying) "fanout gate functions," as well as built-in error correction.

## 3. Digital Encoding in Oligonucleotide Sequences, Error Correction, and Copying

We propose to use short oligonucleotide sequences (about 10 bases to ensure a stable hybridization at room temperature) to encode the digital information. For example, base T will be used to encode "1" and base C will be used to encode "0". An oligonucleotide composed of $n$ ($n = 10$ in the following example) repeated T bases will encode "1" with $n$ repetitions, while an oligonucleotide composed of $n$ repeated C bases will encode "0" with $n$ repetitions. These oligonucleotides could include some other bases that would represent errors in the encoded sequence (e.g., the polyT-nucleotide could include some C, G or A bases as error sites). Magnetic nanoparticles (Au coated, with magnetic cores), synthesized by a known technique[51] and functionalized with single oligonucleotide chains ($A_n$ and $G_n$) complementary to those that are used for encoding of "1" and "0", can be used to recognize the encoded oligonucleotides and to correct errors appearing in the sequences. This is illustrated in Figure 1. Thiol anchor groups at the ends of the oligonucleotides can be used to bind them to the Au shell of the magnetic nanoparticles. Note that here and below, we develop the concepts; details of the experimental procedure are being worked out and will be reported separately as our group advances from the modeling/design to the experimental realization of the proposed systems.

The following reaction steps will be used to recognize the encoded signal, to correct errors in the signal and to amplify it. The present example will demonstrate the error correction process in the poly-T-oligonucleotide (encoded "1" repeated $n - m$ times) with $m$ included errors represented by X (i.e. $m$ foreign bases). A similar process can be used to make error-correction in the poly-C-oligonucleotide (encoded "0" repeated $n - m$ times). Since the encoded signal coming to the system is unknown in advance (it could be poly-T or poly-C), the system is supposed to be ready to read out any incoming signal. Thus the system should include both poly-A- and poly-G-functionalized magnetic nanoparticles. The poly-A-functionalized magnetic nanoparticles will be responsible for hybridization and recognition of poly-T-oligonucleotide, while the poly-G-functionalized magnetic nanoparticles will hybridize with poly-C-oligonucleotide.

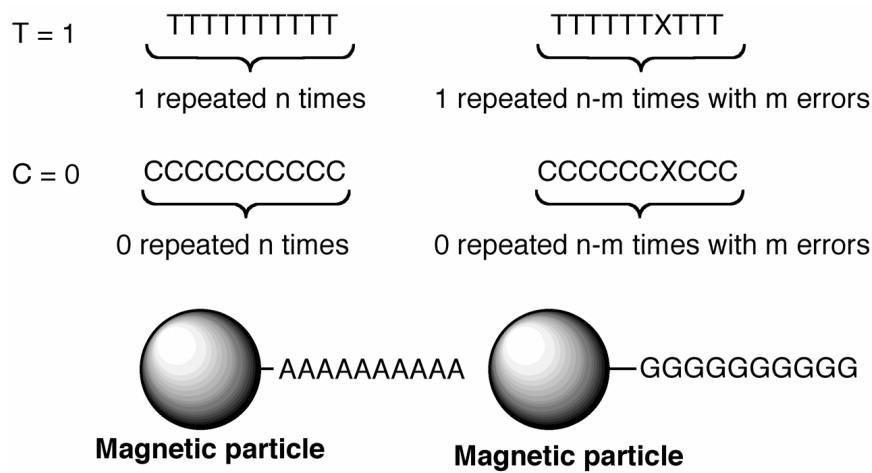

**Figure 1:** Chemical materials used for the error-correction in the encoded oligonucleotides.

Step *a* (Figure 2) shows the hybridization process between poly-T oligonucleotides (incoming signal "1" repeated $n - m$ times with $m$ errors in the sequence) and the complementary poly-A-functionalized magnetic nanoparticles. This process will result in the formation of double-stranded (ds) oligonucleotides (poly-T/poly-A) bound to the magnetic nanoparticles. The hybridization will proceed only if the "melting" point of the ds-oligonucleotide (the temperature corresponding to the dissociation of the ds-DNA complex) is higher than the temperature of the reaction solution. The "melting" point depends on the number of errors in the poly-T sequence — when the number of errors is higher the hybridized complex is less stable and the "melting" point is lower.[52,53] The change of the "melting" point for short ds-oligonucleotide sequences could be ca. 10°C per one error in the sequence. Thus, by selecting different temperatures (e.g., in the range 5-45ºC) for the hybridizing solution, we can control the number of errors in the encoded sequence that will still allow the hybridization process to proceed. The poly-G-functionalized magnetic nanoparticles will stay non-hybridized in the solution. (Note that the magnetic nanoparticles will be applied in the molar excess compared to the encoded oligonucleotides. Thus an excess of the non-hybridized poly-A-functionalized magnetic nanoparticles will also exist in the mixture — not shown in Figure 2).

The reaction mixture composed of magnetic nanoparticles functionalized with the double-stranded and single-stranded oligonucleotides can then be separated (step *b*) on a column modified with poly-C and poly-T oligonucleotides ("DNA-modified column"). All magnetic nanoparticles functionalized with single-stranded (non-hybridized) oligonucleotides will be absorbed by the column due to the hybridization with the complementary strands, and only the magnetic nanoparticles modified with the double-stranded oligonucleotides will go through the column.

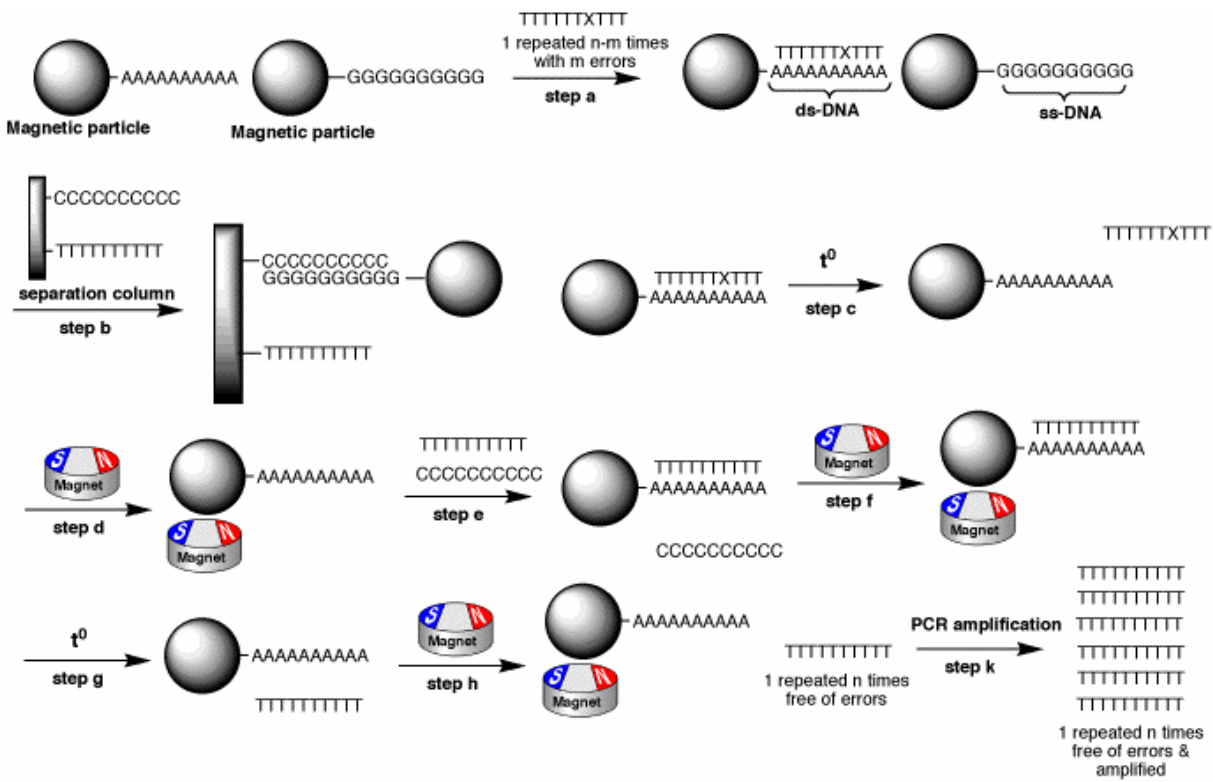

**Figure 2:** Scheme showing the steps of the recognition, error correction and amplification of the encoded DNA signal ($t^0$ denotes heating).

Then the solution of the ds-poly-T/poly-A-functionalized nanoparticles can be heated (e.g., to 80ºC) to dissociate the double-stranded complex (step *c*). This will release the original poly-T oligonucleotide that still contains errors in the sequence and yield the magnetic nanoparticles that are functionalized with the error-free sequence (poly-A) complementary to the encoded oligonucleotide. The magnetic nanoparticles will be separated from the solution using an external magnet, washed from the adsorbed poly-T and re-dispersed in a solution (step *d*).

Since it is not known in advance what kind of the encoded oligonucleotide was added (poly-T or poly-C) and thus it is not known what kind of the functionalized magnetic nanoparticles is obtained at this step, a mixture of poly-T and poly-C oligonucleotides will be added to the solution. It should be noted that these oligonucleotides are different from the original encoded inputs since they are free of errors. A complementary oligonucleotide (poly-T in the present example) will hybridize with the oligonucleotide (poly-A) bound to the magnetic nanoparticles (step *e*). The ds-oligonucleotide-functionalized magnetic nanoparticles will be again collected with the external magnet, washed from the non-hybridized oligonucleotide (poly-C in the present example) and re-dispersed in a solution (step *f*). The temperature will be elevated again (to 80ºC) to dissociate the ds-oligonucleotide, and the magnetic nanoparticles will be collected and separated from the solution with the external magnet (step *h*). The solution will contain the error-corrected equivalent of the initial encoded oligonucleotide (poly-T).

This sample can then be subjected to polymerase chain reaction amplification to generate numerous copies of the error-free encoded oligonucleotide (step *k*). The generated error-free encoded oligonucleotides can be confirmed/analyzed, as well as compared to the original error-containing encoded oligonucleotides by using known DNA-sensing techniques.[54-60]

## 4. Gate Functions

In this section, as a demonstration of our approach we devise two simple biocomputing logic gates for which our error-free encoded oligonucleotides can be used as input signals. The first example, Figure 3, shows the principle of an XOR logic gate based on a DNA hairpin and using poly-T (encoded "1" with *n* repeat units) and poly-C (encoded "0" with *n* repeat units) as input signals. DNA hairpin structures have already been reported as major components of logic gates,[61] however here they will be utilized in a different way. The present example is based on the DNA hairpin that is functionalized with a fluorescent dye (a rhodamine derivative) and a quencher (a quinone molecule) covalently bound to the 5' and 3' ends of the DNA, respectively. The loop part of the DNA contains two domains (poly-A and poly-G) that can hybridize the added encoded poly-T and poly-C oligonucleotides. The sticky ends of the DNA provide the DNA loop formation and they maintain the hairpin structure until the hybridization of the loop part of the DNA substantially rigidifies the DNA molecule.

The length of the DNA loop and the length of the sticky domains can be selected in such a way that the hybridization of only one domain (ether poly-A or poly-G) does not result in the opening of the hairpin structure, while the hybridization of both of them results in the structure that is rigid enough to open the sticky ends and to separate the dye and the quencher. When both inputs (a and b) are the same ("1" encoded by poly-T or "0" encoded by poly-C), the hybridization occurs at one binding site only (domain poly-A or domain poly-G) that does not result in the opening of the DNA hairpin, thus keeping the dye and quencher in close proximity, without fluorescence generation (output signal "0"). If the input signals are different ("0"-"1" or "1"-"0" encoded by poly-T and poly-C oligonucleotides), the hybridization proceeds at both binding domains (poly-A and poly-G), yielding the open structure with a long distance separating the dye and quencher. In this case the generated fluorescence represents the output signal "1". Thus, the output of this gate corresponds to the XOR logic operation.

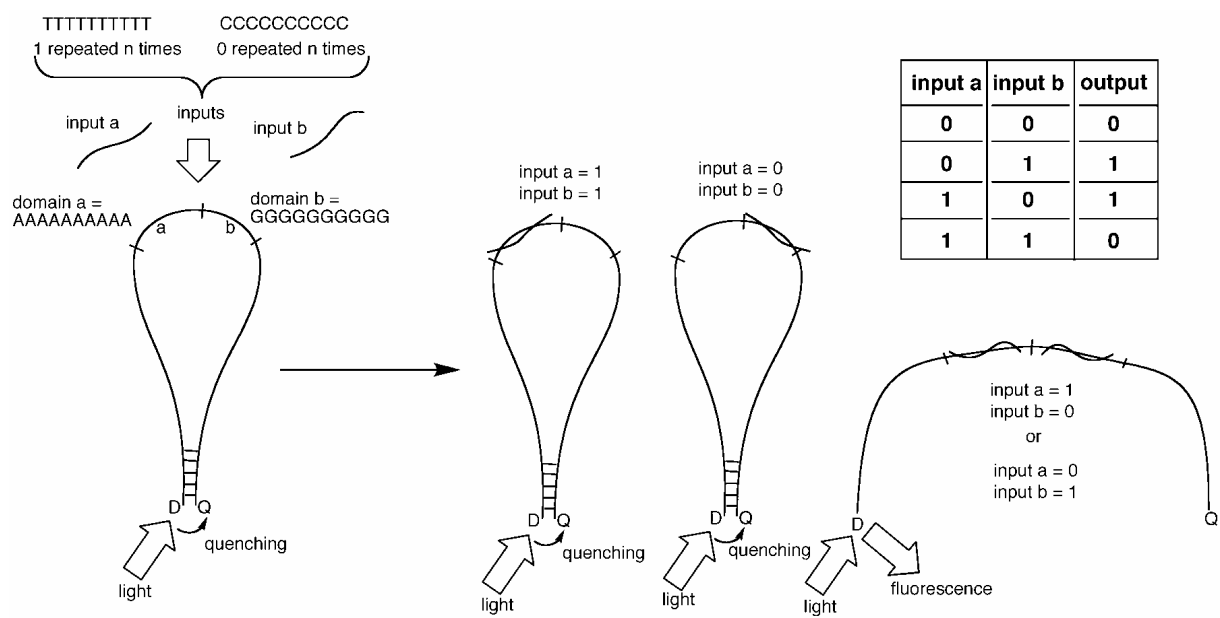

**Figure 3:** Design of an XOR logic gate based on the DNA hairpin structure functionalized with the fluorescent dye (D) and quencher (Q) covalently bound to the 5' and 3' ends, respectively, and using the encoded poly-T and poly-C oligonucleotides ("1" and "0", respectively) as input signals.

The second example, Figure 4, shows a NAND logic gate based on a catalytic DNAzyme and the encoded poly-T and poly-C oligonucleotides ("1" and "0", respectively) as input signals. Catalytic DNAzymes that were discovered recently[22] have already found applications in biosensor systems,[62] however their use in biomolecular logic gates has not yet been experimentally demonstrated. The DNAzyme composed of a folded G-quadruplex cage and the associated heme unit is known to demonstrate a peroxidase-type catalytic activity,[63] as shown in Figure 4. The peroxidase activity can be used to oxidize the NADH cofactor to $NAD^+$ in the presence of $H_2O_2$. The G-quadruplex cage can be hybridized with a complementary poly-C oligonucleotide (encoded "0") resulting in the opening of the cage and the loss of the catalytic activity. This will not happen in the presence of poly-T oligonucleotide (encoded "1"). Thus, in case of any input signal equal to "0", the catalytic activity will be lost and NADH cofactor will not be oxidized.

The output signal of this logic gate is considered to be "1" when the NADH concentration is high (the signal can be detected by the optical absorption at the wavelength 340 nm). The decrease of the NADH concentration in the course of the reaction biocatalyzed by the DNAzyme results in the output signal "0". Any input signal equal to "0" encoded by poly-C oligonucleotide ("0"-"0", "0"-"1", "1"-"0") will result in the deactivation of the catalytic DNAzyme and the preservation of the original concentration of NADH (output signal "1"), while the input signals equal to "1" (encoded by poly-T) ("1"-"1") will preserve the biocatalytic activity of the DNAzyme and reduce the NADH concentration (output signal "0"). The truth table describing the logic function of the gate corresponds to the NAND gate. It should be noted that both exemplified logic gates, Figures 3 and 4, represent universal gates (XOR and NAND) and thus they can be applied to construct logic circuits of any complexity. In addition the NAND gate in Figure 4 has the advantage that the output signal is represented by a concentration of a common biochemical cofactor (NADH) that can be easily coupled to various enzyme-based logic gates.[9–12]

## 5. Signal Rectifier Function for Biochemical Computing

After a successful cycle of DNA information processing one may expect small leftover traces of the initial signal oligonucleotide sequences (encoding "0" or "1"), which for some reasons escaped processing during the cycle. This will result in additional noise in the output signal. Here we devise a contrast enhancement pretreatment that effectively "rectifies" the signal, which is a variant on nonlinear analog signal processing. Just for definiteness, let us assume that the correct signal is "1" and the noisy admixture is a small concentration of "0". (The proposed processing is also effective for the opposite case.)

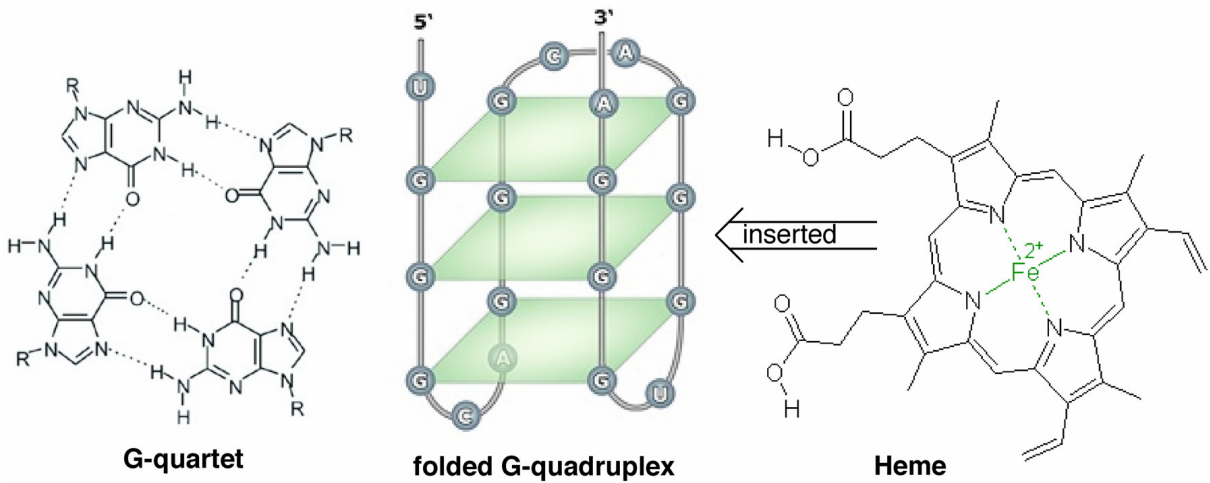

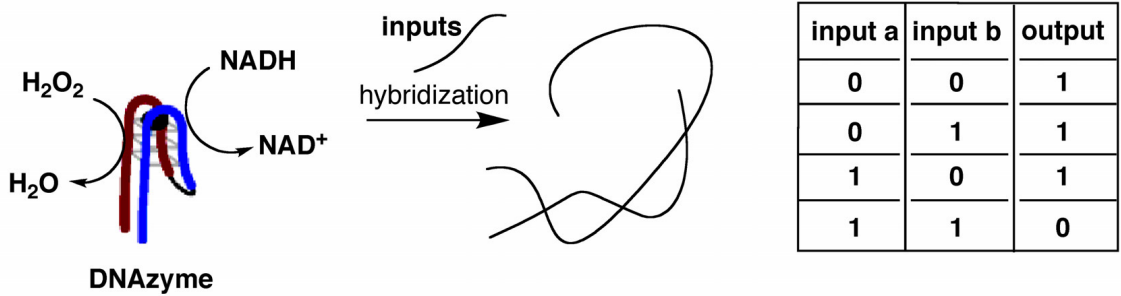

**Figure 4:** Design of a NAND logic gate based on DNAzyme performing the biocatalytic oxidation of NADH by $H_2O_2$, and using the encoded poly-T and poly-C oligonucleotides ("1" and "0", respectively) as input signals.

The initial solution will be composed of a mixture of poly-T oligonucleotide (encoding input signal "1") in concentration C larger than some threshold value $C_T$, and the poly-C oligonucleotide (encoding the noise signal "0") in concentration $C_N$ smaller than threshold value $C_T$. Magnetic nanoparticles functionalized with poly-G oligonucleotide, that is complementary to poly-C, will be added to the solution in concentration $C_T$. In the next step we add the same amount of magnetic nanoparticles functionalized with poly-A oligonucleotide that is complementary to poly-T which is used to encode "1". Magnetic nanoparticles will recognize the encoded oligonucleotides (poly-T and poly-C) and initiate formation of the double-stranded sequences. The concentration of the poly-G-functionalized magnetic nanoparticles is large enough to bind all the noise sequences (poly-C), while the concentration of the poly-A-functionalized magnetic nanoparticles is not sufficient to bind all of the input signal sequences (poly-T). Therefore the resulting solution will represent a mixture of the input signal "1" oligonucleotide sequences (poly-T) at concentration $C - C_T$, magnetic nanoparticles with double-stranded poly-A/poly-T sequences at concentration $C_T$, magnetic nanoparticles with double-stranded poly-G/poly-C sequences at concentration $C_N$, and non-hybridized magnetic nanoparticles with poly-G sequences at concentration $C_T - C_N$.

We next separate the mixture components. By using an external magnet, we will remove from the solution all three types of functionalized magnetic nanoparticles. The solution will now contain only the poly-T oligonucleotides ("1"), but at a somewhat smaller concentration, $C - C_T$. In the last step, the purified input signal (poly-T) will be subject to the polymerase chain reaction amplification to make sure that the concentration of input sequences will be restored to a value above the threshold value $C_T$.

**6. Summary**

In summary, we proposed systems for information processing using encoded DNA sequences, DNAzyme biocatalyzed reactions, and DNA-functionalized magnetic nanoparticles. These systems should allow the first, proof-of-concept realizations of digital biochemical computing, including error correction by utilizing redundancy, as well as two simple logic gates, and "analog" signal rectification. We acknowledge funding by the US-NSF under grant CCF-0726698.

**References**

1.  Unconventional Computing 2005: From Cellular Automata to Wetware, C. Teuscher and A. Adamatzky (Eds.), Luniver Press (2005).
2.  From Utopian to Genuine Unconventional Computers, A. Adamatzky and C. Teuscher (Eds.), Luniver Press (2006).
3.  Y.M. Yin and X. Q. Lin, Progress in Chemistry 13, 337-342 (2001).
4.  A. Adamatzky, B. De Lacy Costello and T. Asai, Reaction-Diffusion Computers, Elsevier Science (2005).
5.  A. Adamatzky, Computing in Nonlinear Media and Automata Collectives, Taylor & Francis (2001).
6.  Molecular Computing, T. Sienko, A. Adamatzky, N. G. Rambidi and M. Conrad (Eds.), MIT Press (2005).
7.  A. Adamatzky, IEICE Trans. Electronics E87C, 1748-1756 (2004).
8.  G. Bell and J.N. Gray, in: Beyond Calculation: The Next Fifty Years of Computing, P. J. Denning and R.M. Metcalfe (Eds.), Chap. 1, p. 30, Copernicus/Springer (1997).
9.  R. Baron, O. Lioubashevski, E. Katz, T. Niazov and I. Willner, Angew. Chem. 45, 1572-1576 (2006).

10. R. Baron, O. Lioubashevski, E. Katz, T. Niazov and I. Willner, Org. Biomol. Chem. 4, 989-991 (2006).
11. R. Baron, O. Lioubashevski, E. Katz, T. Niazov and I. Willner, J. Phys. Chem. A 110, 8451-8456 (2006).
12. T. Niazov, R. Baron, E. Katz, O. Lioubashevski and I. Willner, Proc. Natl. Acad. USA 103, 17160-17163 (2006).
13. L. M. Adleman, Science 266, 1021-1024 (1994).
14. M. Amos, Theoretical and Experimental DNA Computation, Springer (2005).
15. G. Paun, G. Rozenberg and A. Salomaa, DNA Computing - New Computing Paradigms, Springer (1998).
16. M. N. Stojanovic, D. Stefanovic, T. LeBean and H. Yan, in: Bioelectronics: From Theory to Applications, I. Willner and E. Katz (Eds.), Wiley-VCH, Chap. 14, pp. 427-455 (2005).
17. Z. Ezziane, Nanotechnology 17, R27-R39 (2006).
18. N. Jonoska, J. Computer Sci. Techn. 19, 98-113 (2004).
19. L. Kari, G. Gloor and S. Yu, Theoretical Computer Science 231, 192-203 (2000).
20. J. Watada, S. Kojima, S. Ueda and O. Ono, Int. J. Innov. Comp. Inf. 2, 273-282 (2006).
21. A. L. Han and D.M. Zhu, Computational Intelligence and Bioinformatics, Pt 3, Proc. Lecture Notes in Computer Science 4115, 328-335 (2006).
22. J.C. Achenbach, W. Chiuman, R.P.G. Cruz and Y. Li, Curr. Pharm. Biotechnol. 5, 321-336 (2004).
23. Y.F. Wang, G.Z. Cui, B.Y. Huang, L.Q. Pan and X.C. Zhang, Computational Intelligence and Bioinformatics, Pt 3, Proceedings Lecture Notes in Computer Science 4115, 248-257 (2006).
24. Z.Z. Zhang, C.H. Fan and L. He, Current Nanoscience 1, 91-95 (2005).
25. D. van Noort and L.F. Landweber, DNA Computing Lecture Notes in Computer Science 2943, 190-196 (2004).
26. I. Willner, E. Katz, B. Willner, R. Blonder, V. Heleg-Shabtai and A.F. Bückmann, Biosens. Bioelectron. 12, 337-356 (1997).
27. I. Willner and E. Katz, Angew. Chem. 39, 1180-1218 (2000).
28. A.P. de Silva, Nat. Mater. 4, 15-16 (2005).
29. C.P. Collier, E.W. Wong, M. Belohradsky, F.M. Raymo, J.F. Stoddart, P.J. Kuekes, R.S. Williams and J.R. Heath, Science 285, 391-394 (1999).
30. A. Adamatzky and B. De Lacy Costello, Phys. Rev. E 66, 046112 (2002).
31. A.P. de Silva and N.D. McClenaghan, Chem. Eur. J. 8, 4935-4945 (2002).
32. A.N. Shipway and I. Willner, Acc. Chem. Res. 34, 421-432 (2001).
33. A. Credi, V. Balzani, S.J. Langford, and J.F. Stoddart, J. Am. Chem. Soc. 119, 2679-2681 (1997).

34. A.P. de Silva, H.Q.N. Gunaratne and C.P. McCoy, J. Am. Chem. Soc. 119, 7891-7892 (1997).

35. J.F. Callan, A.P. de Silva and N.D. McClenaghan, Chem. Commun., 2048-2049 (2004).

36. A. Doron, M. Portnoy, M. Lion-Dagan, E. Katz and I. Willner, J. Am. Chem. Soc. 118, 8937-8944 (1996).

37. T. Gunnlaugsson, D.A. MacDonail and D. Parker, Chem. Commun. 93-94 (2000).

38. F. Li, M. Shi, C. Huang and L. Jin, J. Mater. Chem. 15, 3015-3020 (2005).

39. A.P. de Silva and N.D. McClenaghan, J. Am. Chem. Soc. 122, 3965-3966 (2000).

40. J. Andreasson, G. Kodis, Y. Terazono, P. A. Liddell, S. Bandyopadhyay, R.H. Mitchell, T.A. Moore, A.L. Moore and D. Gust, J. Am. Chem. Soc. 126, 15926-15927 (2004).

41. D.H. Qu, Q.C. Wang and H. Tian, Angew. Chem. 44, 5296-5299 (2005).

42. S.J. Langford and T. Yann, J. Am. Chem. Soc. 125, 11198-11199 (2003).

43. D. Margulies, G. Melman and A. Shanzer, Nat. Mater. 4, 768-771 (2005).

44. X.P. Su and L.M. Smith, Nucleic Acids Res. 32, 3115-3123 (2004).

45. J. Macdonald, Y. Li, M. Sutovic, H. Lederman, K. Pendri, W.H. Lu, B.L. Andrews, D. Stefanovic and M.N. Stojanovic, Nano Lett. 6, 2598-2603 (2006).

46. E. Katz, A.F. Bückmann and I. Willner, J. Am. Chem. Soc. 123, 10752-10753 (2001).

47. R. Unger and J. Moult, Proteins 63, 53-64 (2006).

48. J. Liu, D.P. Wernette and Y. Lu, Angew. Chem. 44, 7290-7293 (2005).

49. O. Pretzel, Error-Correcting Codes and Finite Fields, Oxford University Press (1996).

50. B.P. Lathi, Linear Systems and Signals, Oxford University Press (2005).

51. M. Pita, A. Ballesteros, C. Vaz, J.M. Abad, C. Briones, E. Mateo-Martí, J.A. Martín-Gago, M.P. Morales and V.M. Fernández, submitted for publication (2007).

52. J.B. Fiche, A. Buhot, R. Calemczuk and T. Livache, Biophys. J. 92, 935-946 (2007).

53. N.C. Harris and C.H. Kiang, J. Phys. Chem. B 110, 16393-16396 (2006).

54. I. Willner, E. Katz and B. Willner, in: Sensors Updates, Vol. 5, H. Baltes, W. Göpel and J. Hesse (Eds.), Wiley-VCH, Chap. 2, pp. 45-102 (1999).

55. I. Willner, E. Katz and B. Willner, in: Electroanalytical Methods of Biological Materials, A. Brajter-Toth and J.Q. Chambers (Eds.), Marcel Dekker, pp. 43-107 (2002).

56. E. Katz and I. Willner, in: Ultrathin Electrochemical Chemo- and Biosensors. Technology and Performance, V. Mirsky (Ed.), Springer Series on Chemical Sensors and Biosensors. Methods and Applications, O. Wolfbeis (Series Ed.), Springer-Verlag, Chap. 4, pp. 67-106 (2004).

57. E. Katz, in: Smart Sensors and MEMS, S.Y. Yurish and M.T.S.R. Gomes (Eds.), Springer Verlag, NATO Science Series, Vol. 181, Chap. 14, pp. 447-474 (2005).

58. E. Katz, B. Willner and I. Willner, in: Perspectives in bioanalysis, Vol. 1, Electrochemistry of nucleic acids and proteins – Towards electrochemical sensors for genomics and proteomics, E. Palecek, F. Scheller and J. Wang (Eds.), Elsevier, Amsterdam, pp. 195-246 (2005).

59. E. Katz and I. Willner, Electroanalysis 15, 913-947 (2003).

60. F. Patolsky, Y. Weizmann, E. Katz and I. Willner, Angew. Chem. 42, 2372-2376 (2003).

61. Z.X. Yin, J. Z. Cui, J. Yang and J. Xu, Computational Intelligence and Bioinformatics, Pt 3, Proc. Lecture Notes in Computer Science 4115, 238-247 (2006).

62. J.W. Liu and Y. Lu, Chem. Mater. 16, 3231-3238 (2004).

63. P. Travascio, Y.F. Li and D. Sen, Chemistry & Biology 5, 505-517 (1998).